\documentclass[a4paper,american]{scrartcl}
\usepackage{lmodern}
\usepackage[utf8]{inputenc}
\usepackage[T1]{fontenc}
\usepackage{color}
\usepackage{todonotes}
\usepackage{babel}
\usepackage{amsmath}
\usepackage{amssymb}
\usepackage{setspace}
\usepackage[authoryear]{natbib}
\usepackage[font=small]{quoting}
\usepackage{latexsym}
\usepackage[bookmarks=false,pdfborder={0 0 0},colorlinks=false]{hyperref}

\makeatletter

\newcounter{contatore}
\newcounter{contatore1}
\newcounter{contatore2}
\makeatother

\begin{document}

\title{Individuality and the account of non-locality: the case for the particle ontology in quantum physics}

\author{Michael Esfeld\thanks{University of Lausanne, Department of Philosophy,
    CH-1015 Lausanne, Switzerland. E-mail:
    \protect\href{mailto:michael-andreas.esfeld@unil.ch}{michael-andreas.esfeld@unil.ch}}
}
\maketitle
\begin{abstract}
      \begin{center} For Olimpia Lombardi (ed.), \emph{Quantum worlds: identity, indistinguishability and non-locality in quantum physics},
      
      Cambridge University Press, forthcoming   \end{center}
    \medskip

\noindent The paper explains why an ontology of permanent point particles that are individuated by their relative positions and that move on continuous trajectories as given by a deterministic law of motion constitutes the best solution to the measurement problem in both quantum mechanics and quantum field theory. This case is made by comparing the Bohmian theory to collapse theories such as the GRW matter density and the GRW flash theory. It is argued that the Bohmian theory makes the minimal changes, concerning only the dynamics and neither the ontology nor the account of probabilities, that are necessary to get from classical mechanics to quantum physics. There is no cogent reason to go beyond these minimal changes. 
\medskip{}

\end{abstract}

\tableofcontents{}

\section{The measurement problem}
Quantum mechanics is a highly successful theory as far as the prediction, confirmation and application of measurement outcome statistics is concerned. The central tool for these predictions is Born's rule according to which, in brief, the squared modulus of the wave function $\psi$ of a quantum system indicates the probability to find a particle at a certain location if a measurement is made. Consequently, the measurement outcomes show a $|\psi|^2$ distribution. Furthermore, all the information that we obtain in experiments is knowledge about positions, as \citet[][p. 166]{Bell:2004aa} stressed: the measurement outcomes are recorded in macroscopic positions -- such as the positions of dots on a screen, pointer positions, etc. and finally brain configurations --, and they provide information about where the investigated objects are. This insight holds whatever observables one defines and measures in terms of operators. Thus, for instance, all the information that the outcome of a spin measurement of an electron by means of a Stern-Gerlach experiment provides is information about the wave packet in which the electron is located, etc.

However, an algorithm to calculate measurement outcome statistics is not a physical theory. Physics is about nature, \emph{physis} in ancient Greek. Consequently, a physical theory has (i) to spell out an ontology of what there is in nature according to the theory, (ii) to provide a dynamics for the elements of the ontology and (iii) to deduce measurement outcome statistics from the ontology and dynamics by treating measurement interactions within the ontology and dynamics; in order to do so, the ontology and dynamics have to be linked with an appropriate probability measure. Thus, the question is what is the law that describes the individual processes that occur in nature (dynamics) and what are the entities that make up these individual processes (ontology).

Consider as an illustration the double slit experiment in quantum mechanics: one can do this experiment with individual particles, say prepare a source that emits one particle every morning at 8 a.m. so that one gets an outcome recorded in the form of exactly one dot on a screen every morning. The question then is what happens between the emission of the particle from the source and the measurement record on the screen. Is there a particle that goes through one of the two slits? A wave that goes through both slits and that contracts afterwards to be recorded as a dot on a screen? Or something else? The constraint on the ontology and dynamics that are to answer this question is that they have to account for the characteristic distribution of the dots on the screen that shows up if one runs this experiment many times. In other words, the ontology and the dynamics have to explain the measurement outcome distribution.

It is not possible to infer the law that describes the individual processes that occur in nature from the rule to calculate the measurement outcome statistics. The characteristic pattern of the measurement outcome distribution in the double slit experiment by no means reveals what happens between the source and the screen. Trying to make such inferences runs into the famous measurement problem of quantum physics.  The measurement outcomes show a $|\psi|^2$ distribution, the law for the evolution of the wave function $\psi$ is the Schrödinger equation, but the  Schrödinger evolution does in general not lead to measurement outcomes. More precisely, the by now standard formulation of the measurement problem is the one of \citet[][p. 7]{Maudlin:1995aa}: \begin{quote}
1.A The wave-function of a system is \emph{complete}, i.e. the wave-function specifies (directly or indirectly) all of the physical properties of a system.

\noindent 2.A The wave-function always evolves in accord with a linear dynamical equation (e.g. the Schrödinger equation).

\noindent 3.A Measurements of, e.g., the spin of an electron always (or at least usually) have determinate outcomes, i.e., at the end of the measurement the measuring device is either in a state which indicates spin up (and not down) or spin down (and not up).
\end{quote}
      
\noindent Any two of these propositions are consistent with one another, but the conjunction of all three of them is inconsistent. This can be easily illustrated by Schrödinger's cat paradox (\cite{Schroedinger:1935aa}, p. 812): if the entire system is completely described by the wave function and if the wave function always evolves according to the Schrödinger equation, then, due to the linearity of this wave equation, superpositions and entangled states will in general be preserved. Consequently, a measurement of the cat will in general not have a determinate outcome: at the end of the measurement, the cat will not be in the state of either being alive or being dead.

Hence, the measurement problem is not just a -- philosophical -- problem of the interpretation of a given formalism. It concerns the very formulation of a consistent quantum theory. Even if one takes (1.A) and (2.A) to define the core formalism of quantum mechanics and abandons (3.A), one has to put forward a formulation of quantum physics that establishes a link with at least the appearance of determinate measurement outcomes. If one retains (3.A), one has to develop a formulation of a quantum theory that goes beyond a theory in which only a wave function and a linear dynamical equation for the evolution of the wave function figure. Accordingly, the solution space for the formulation of a consistent quantum theory can be divided into many worlds theories, rejecting (3.A), collapse theories, rejecting (2.A), and additional variable theories, rejecting (1.A).

However, research in the last decade has made clear that we do not face three equally distinct possibilities to solve the measurement problem, but just two: the main dividing line is between endorsing (3.A) and rejecting it. If one endorses (3.A), the consequence is not that one has to abandon \emph{either} (1.A) \emph{or} (2.A), but that one has to amend \emph{both} (1.A) and (2.A). Determinate measurement outcomes as described in (3.A) are outcomes occurring in ordinary physical space, that is, in three-dimensional space or four-dimensional space-time. Hence, endorsing (3.A) entails being committed to the existence of a determinate configuration of matter in physical space that constitutes measurement outcomes (such as a live cat, or an apparatus configuration that indicates spin up, etc.). If one does so, one cannot stop at amending (2.A). The central issue then is not whether or not a collapse term for the wave function has to be added to the Schrödinger equation, because even with the addition of such a term, this equation still is an equation for the evolution of the wave function, by contrast to an equation for the evolution of a configuration of matter in physical space. Consequently, over and above the Schrödinger equation -- however amended -- a law or rule is called for that establishes an explicit link between the wave function and the configuration of matter in physical space. By the same token, (1.A) has to be changed in such a way that reference is made to the configuration of matter in physical space and not just the quantum state as encoded in the wave function (see \cite{Allori:2008aa}). This fact underlines the point made above: we need both a dynamics (filling in proposition (2.A) above) and an ontology (filling in proposition (1.A) above) that specifies the entities to which the dynamics refers and whose evolution it describes. The way in which these two interplay then has to account for the measurement outcomes and their distribution (filling in proposition (3.A) above). 

This point can be further illustrated by a second formulation of the measurement problem that \citet[][p. 11]{Maudlin:1995aa} provides: \begin{quote}
1.B The wave-function of a system is \emph{complete}, i.e. the wave-function specifies (directly or indirectly) all of the physical properties of a system.

\noindent 2.B The wave-function always evolves in accord with a deterministic dynamical equation (e.g. the Schrödinger equation).

\noindent 3.B Measurement situations which are described by identical initial wave-functions sometimes have different outcomes, and the probability of each possible outcome is given (at least approximately) by Born's rule.
\end{quote}
 
\noindent Again, any two of these propositions are consistent with one another, but the conjunction of all three of them is inconsistent. Again, the issue is what the law is and what the physical entities are to which the law refers such that, if one takes (3.B) for granted, configurations of matter in physical space that constitute definite measurement outcomes are accounted for.

All mathematical formulations of non-relativistic quantum mechanics work with a formalism in terms of a definite, finite number of point particles and a wave function that is attributed to these particles, with the basic law for the evolution of the wave function being the Schrödinger equation. The wave function is defined on configuration space, thereby taking for granted that the particles have a position in three-dimensional space: for $N$ particles, the configuration space has $3N$ dimensions so that each point of configuration space represents a possible configuration of the $N$ particles in three-dimensional space. This fact speaks also in quantum physics against regarding configuration space as the physical space, since its dimension depends on a definite number of particles admitted in three-dimensional space.

Even if one pursues an ontology of configuration space being the physical space in quantum physics, one can add further stuff than the wave function to configuration space -- such as e.g. the position of a world-particle in configuration space -- or take the wave function to collapse occasionally in configuration space in order to solve the measurement problem (see \cite{Albert:2015aa}, chs. 6-8). However, the problem remains how to connect what there is in configuration space and its evolution with our experience of three-dimensional physical objects, their relative positions and motions. That experience is the main reason to retain (3.A and B).

This paper is situated in the framework that envisages abandoning (3.A and B) only as a last resort, that is, only in case it turned out that the consequences of the options that endorse (3.A and B) were even more unpalatable than the ones of rejecting (3.A and B). It seeks to make a contribution to assessing the options that are available in this framework, namely the option that starts from amending the Schrödinger dynamics by admitting a dynamics of the collapse of the wave function (not 2.A and B) (next section) and the option that starts from admitting particles, although the information about their positions is not contained in the wave function (not 1.A and B) (section 3). The paper closes with a few remarks on quantum field theory (section 4).

\section{The collapse solution and its ontology}
Against this background, let us take (3.A and B) for granted. On the one hand, the fact that all formulations of non-relativistic quantum mechanics work with a formalism in terms of a definite, finite number of point particles and a wave function that is attributed to these particles suggests to propose a particle ontology for quantum physics. Following this suggestion, the basic ontology -- that is, the objects in nature to which the formalism refers -- is the same in classical and quantum mechanics. By contrast, the dynamics that is postulated for these objects is radically different: there is no wave function in classical mechanics. This view is supported by the fact that, as mentioned in the previous section, all recorded measurement outcomes consist in definite positions of macroscopic, discrete objects that provide information about where microscopic, discrete objects (i.e. particles) are located. Furthermore, the measurement instruments are composed of particles that hence are located where these instruments are.

On the other hand, the dynamics as given by the wave function evolving according to the Schrödinger equation does not describe the evolution of particle positions: it does not provide for determinate trajectories of individual particles. Even if one starts with an initial condition of precise information about particle positions, the Schrödinger equation will in general describe these particles as evolving into a superposition of different trajectories. Consequently, the Schrödinger evolution does not establish an intertemporal identity of these objects: it fails to distinguish them. Moreover, the Heisenberg uncertainty relations put a limit on the epistemic accessibility of particle positions: operators for position and momentum cannot both be measured with arbitrary accuracy on a quantum system. These facts motivate going for another quantum ontology than the one of particles. In brief, if there are no precise particle positions, it makes no sense to maintain a particle ontology. An object that does not have a precise position is not a particle, but something else. The foremost candidate for that something else is a wave, since the Schrödinger equation is a wave equation.

There is a proposal for a quantum ontology that takes the wave equation to describe a wave in three-dimensional, physical space, namely a continuous matter density field (see \cite{Ghirardi:1995aa}). Determinate measurement outcomes -- as well as the formation of discrete objects in general -- are accounted for in terms of a spontaneous contraction of the matter density field at certain points or regions of space. This spontaneous contraction is represented in terms of the collapse of the wave function. Consequently, \cite{Ghirardi:1995aa} use a modified Schrödinger dynamics that breaks the linearity and the determinism of the Schrödinger equation by including the collapse of the wave function under certain circumstances (rejection of (2.A and B)).

On the quantum dynamics proposed by Ghirardi, Rimini and Weber (GRW) (see \cite{Ghirardi:1986aa}), the wave function undergoes spontaneous jumps in configuration space at random times distributed according to the Poisson distribution with rate $N\lambda$, with $N$ being the particle number and $\lambda$ being the mean collapse rate. Between two successive jumps the wave function $\Psi_t$ evolves according to the usual Schrödinger equation. At the time of a jump the $k$th component of the wave function $\Psi_t$ undergoes an instantaneous collapse according to
\begin{align}
    \Psi_t(x_1,\dots,x_k,\dots,x_N) \mapsto \frac{(L^x_{x_k})^{1/2}
\Psi_t(x_1,\dots,x_k,\dots,x_N)}{\|(L^x_{x_k})^{1/2} \Psi_t\|}, 
\end{align}
where the localization operator $L^x_{x_k}$ is given as a multiplication operator of the form
\begin{align}
    L^x_{x_k} := \frac{1}{(2\pi\sigma^2)^{3/2}} e^{-\frac{1}{2\sigma^2}(x_k-x)^2},
\end{align}

\noindent and $x$, the centre of the collapse, is a random position distributed according to the probability density $p(x)=\|(L^x_{x_k})^{1/2} \Psi_t\|^2$. This modified Schrödinger evolution captures in a mathematically precise way what the collapse postulate in the textbooks introduces by a \emph{fiat}, namely the collapse of the wave function so that it can represent localized objects in physical space, including in particular measurement outcomes. GRW thereby introduces two additional parameters, the mean rate $\lambda$ as well as the width $\sigma$ of the localization operator. An accepted value of $\lambda$ is of the order of $10^{15}s^{-1}$. This value implies that the spontaneous localization process for a single particle occurs only at astronomical time scales of the order of $10^{15}s$, while for a macroscopic system of $N\sim10^{23}$ particles, the collapse happens so fast that possible superpositions are resolved long before they would be experimentally observable. Moreover, the value of $\sigma$ can be regarded as localization width; an accepted value is of the order of $10^{-7}m$.

A further law then is needed to link this modified Schrödinger equation with a wave ontology in the guise of an ontology of a continuous matter density field $m_t(x)$ in space:
\begin{align}
    m_t(x) = \sum_{k=1}^N m_k \int d^3x_1\dots d^3x_N\, \delta^3(x-x_k)
    |\Psi_t(x_1,\dots,x_N)|^2.
\end{align}
This field $m_t(x)$ is to be understood as the density of matter in three-dimensional physical space at time $t$ (see \cite{Allori:2008aa}, section 3.1). The thus defined theory of a GRW collapse dynamics describing the evolution of a matter density field in physical space is known as GRWm.

Consequently, although GRWm is formulated in terms of particle numbers, there are no particles in the ontology. More generally speaking, there is no plurality of fundamental physical systems. There is just one object in the universe, namely a matter density field that stretches out throughout space and that has varying degrees of density at different points of space, with these degrees changing in time. Hence, there are no individual systems in nature according to this theory so that the issue of identity and distinguishability of individual quantum systems does not arise. That notwithstanding, this theory accounts for measurement outcomes that appear as individual particle outcomes in terms of a spontaneous contraction of the matter density field at certain locations. Position thus is distinguished in the form of degrees of matter density at points of space. It is the only fundamental physical property, as \cite{Allori:2013aa} point out: \begin{quote}
Moreover, the matter that we postulate in GRWm and whose density is given by the $m$ function does not ipso facto have any such properties as mass or charge; it can only assume various levels of density. (\cite{Allori:2013aa}, pp. 331--332)
\end{quote}
\noindent This ontology is not committed to a dualism of an absolute space and time on the one hand and a matter field that fills that space and that develops in time on the other hand. It can also be conceived as a relationalism about space and time, namely a field relationism (see \cite{Rovelli:1997aa} for a field relationalism in the context of general relativity theory): there just is the matter density field as an autonomous entity (substance), with an internal differentiation into various degrees of density and change of these degrees. The geometry of a three-dimensional, Euclidean space then is a means to represent that differentiation, and time is a means to represent that change. In any case, on this view, matter is a continuous stuff, known as gunk, and it is a primitive stuff or bare substratum that, moreover, admits of different degrees of density as a primitive matter of fact. There is nothing that accounts for the difference in degrees of density of matter in different regions of space as expressed by the $m$ function in the formalism.

Making the collapse postulate of textbook quantum mechanics precise by amending the Schrödinger equation with the two new parameters $\lambda$ and $\sigma$, indicating the mean rate of spontaneous collapse and the width of the localization operator, paradoxically has the consequence that the GRW formalism cannot reproduce the predictions that textbook quantum mechanics achieves by applying the Born rule in all situations. Rather than being a decisive drawback, this, however, opens up the way for testing collapse theories like GRW against theories that solve the measurement problem without the collapse postulate (see \cite{Curceanu:2016aa} for such experiments).

On a more fundamental level, however, it is in dispute whether the ontology of a continuous matter density field that develops according to the GRW equation is sufficient to solve the measurement problem. The reason is the so-called problem of the tails of the wave function. This problem arises from the fact that the GRW formalism mathematically implements the collapse postulate by multiplying the wave function with a Gaussian, such that the collapsed wave function, although being sharply peaked in a small region of configuration space, does not actually vanish outside that region; it has tails spreading to infinity. On this basis, one can object that GRWm does not achieve its aim, namely to describe measurement outcomes in the form of macrophysical objects having a definite position (see e.g. \citet[][pp. 135-138]{Maudlin:2010aa}). However, one can also make a case for the view that this mathematical fact does not prevent GRWm from accounting for definite measurement outcomes in physical space (see e.g. \cite{Wallace:2014aa} and \citet[][section 3]{Egg:2015aa}).

The main drawback of GRWm arguably is its account of quantum non-locality, which occurs when the wave function collapses all over space. Consider a simple example, namely the thought experiment of one particle in a box that Einstein presented at the Solvay conference in 1927 (the following presentation is based on de Broglie’s version of the thought experiment in \citet[][pp. 28-29]{Broglie:1964aa} and on \cite{Norsen:2005aa}): the box is split in two halves which are sent in opposite directions, say from Brussels to Paris and Tokyo. When the half-box arriving in Tokyo is opened and found to be empty, there is on all accounts of quantum mechanics that acknowledge that measurements have outcomes a fact that the particle is in the half-box in Paris.

On GRWm, the particle is a matter density field that stretches over the whole box and that is split in two halves of equal density when the box is split, these matter densities travelling in opposite directions. Upon interaction with a measurement device, one of these matter densities (the one in Tokyo in the example given above) vanishes, while the matter density in the other half-box (the one in Paris) increases so that the whole matter is concentrated in one of the half-boxes. One might be tempted to say that some matter travels from Tokyo to Paris; however, since it is impossible to assign any finite velocity to this travel, the use of the term ``travel'' is inappropriate. For lack of a better expression let us say that some matter is delocated from Tokyo to Paris (this expression was proposed by Matthias Egg, see \citet[][p. 193]{Egg:2014aa}); for even if the spontaneous localization of the wave function is conceived as a continuous process as in \cite{Ghirardi:1990aa}, the time it takes for the matter density to disappear in one place and to reappear in another place does not depend on the distance between the two places. This delocation of matter, which is not a travel with any finite velocity, is quite a mysterious process that the GRWm ontology asks us to countenance.

Apart from the matter density ontology, there is another ontology for the GRW collapse formalism available. This ontology goes back to \citet[][ch. 22, originally published 1987]{Bell:2004aa}: whenever there is a spontaneous localization of the wave function in configuration space, that development of the wave function in configuration space represents an event occurring at a point in physical space. These point-events are known as flashes; the term ``flash'' was coined by \citet[][p. 826]{Tumulka:2006aa}. According to the GRW flash theory (GRWf), the flashes are all there is in physical space. Macroscopic objects are, in the terms of \citet[][p. 205]{Bell:2004aa}, galaxies of such flashes. Consequently, the temporal development of the wave function in configuration space does not represent the distribution of matter in physical space. It represents the objective probabilities for the occurrence of further flashes, given an initial configuration of flashes. Hence, space is not filled with persisting objects such as particles or fields. There only is a sparse distribution of single events. These events are individuals and distinguishable, because they are absolutely discernible by their position in space; but there is no intertemporal identity of anything, because these events are ephemeral.

GRWf is committed to absolute space and time (or space-time) as the substance in wish the flashes occur. There can also be times at which there are no flashes at all. The flashes, again, are bare particulars. There is no informative answer to the question of what distinguishes an empty space-time point from a point at which a flash occurs: it is a primitive matter of fact that there are flashes at some points of space-time. The flashes are only characterized by their space-time location. They come into existence at some points of space-time out of nothing and they disappear into nothing.

The most serious drawback of GRWf is that this theory covers only the spontaneous appearance and disappearance of flashes, but offers no account of interactions, given that there are no persisting objects at all. The idea that motivates the GRW collapse dynamics is that a macroscopic object such as a measurement device consists of a great number of particles so that the entanglement of the wave function of the apparatus with the one of the measured quantum object will be immediately reduced due to the spontaneous localization of the wave function of the apparatus. However, even if one supposes that a macroscopic object such as a measurement apparatus can be conceived as a galaxy of flashes (but see the reservations of \cite{Maudlin:2011aa}, pp. 257-258), there is on GRWf nothing with which the apparatus could interact: there is no particle that enters it, no matter density and in general no field that gets in touch with it either (even if one conceives the wave function as a field, it is a field in configuration space and not a field in physical space). There only is one flash (standing for what is usually supposed to be a quantum object) in its past light cone, but there is nothing left of that flash with which the apparatus could interact. In brief, there simply are no objects that could interact in GRWf. 

\section{The particle solution and its dynamics}
One may regard the flashes as particles that are deprived of their trajectories, so that there are only disconnected particle localisations left, being represented by the collapses of the wave function in GRWf -- although, of course, particles without trajectories and only disconnected point-like localisation events no longer is a particle ontology. Let us therefore now consider a particle ontology for quantum mechanics. The quantum theory in that sense is the one going back to  \cite{Broglie:1928aa} and \cite{Bohm:1952}. Its dominant contemporary version is known as Bohmian mechanics (BM) (see \cite{Durr:2013aa}). BM is based on the following four axioms:    
\begin{enumerate}
\item \emph{Particle configuration}: There always is a configuration of $N$ permanent point particles in the universe that are characterized only by their positions $x_1,\dots,x_N$ in three-dimensional, physical space at any time $t$.
\item  \emph{Guiding equation}: A wave function is attributed to the particle configuration, being the central dynamical parameter for its evolution. The wave function has the task to determine a velocity field along which the particles move, given their positions. It accomplishes this task by figuring in the law of motion of the particles, which is known as the guiding equation:
\begin{equation}\label{eq:standard-guiding-equation}
\frac{\mathrm{d}x_k}{\mathrm{d}t}=\frac{\hbar}{m_k}\mathrm{Im}\frac{\nabla_k \psi}{\psi} (x_1,\dots,x_N).
\end{equation}
This equation yields the evolution of the $k$th particle at a time $t$ as depending on, via the wave function, the position of all the other particles at that time.
\item \emph{Schrödinger equation}: The wave function always evolves according to the usual Schrödinger equation.
\item \emph{Typicality measure}: On the basis of the universal wave function $\Psi$, a typicality measure can be defined in terms of the $|\Psi|^2$--density. Given that typicality measure, it can then be shown that for almost all initial conditions, the distribution of particle configurations in an ensemble of sub-systems of the universe that admit of a wave function $\psi$ of their own (known as effective wave function) is a $|\psi|^2$--distribution. A universe in which this distribution of the particles in sub-configurations obtains is considered to be in quantum equilibrium.
\end{enumerate}

\noindent Assuming that the actual universe is a typical Bohmian universe in that it is in quantum equilibrium, one can hence deduce Born's rule for the calculation of measurement outcome statistics on sub-systems of the universe in BM (instead of simply stipulating that rule). In a nutshell, the axiom of $|\Psi|^2$ providing a typicality measure with $\Psi$ being the universal wave function justifies applying the $|\psi|^2$--rule for the calculation of the probabilities of measurement outcomes on particular systems within the universe, with $\psi$ being the effective wave function of the particular systems in question (see \citet[][ch. 2]{Durr:2013aa}). 

Axiom (1) defines the ontology of the theory. The universe is a configuration of point particles that, consequently, always have a precise position relative to one another: they stand in determinate distance relations to each other. Indeed, BM does not require the commitment to an absolute space in which the particles are embedded and an absolute time in which their configuration evolves. The geometry and the time with its metric can be conceived as a mere means to represent the particle configuration and its change, that is, the change in the relative distances of the particles (see \citet[][ch. 3.2]{Esfeld:2017aa} for the philosophical argument and \cite{Vassallo:2016aa} for a relationalist formulation of BM). Consequently, the particles are individuals that are absolutely discernible by their position in a configuration. They have an identity in time that is provided by the continuous trajectory that their motion -- i.e. the change in their relative positions -- traces out. In the framework of relationalism about space and time, one can employ the distance to individuate the particles, so that the particles in BM are not bare particulars: the distance relations make them the objects that they are and account for their numerical plurality (see \citet[][ch. 2.1]{Esfeld:2017aa}).

This ontology contains the core of the solution to the measurement problem that BM provides: there always are point particles with definite positions, and these particles compose the macroscopic objects. Hence, Schrödinger's cat is always either alive or dead, a radioactive atom is always either decayed or not decayed, an electron in the double slit experiment with both slides open always goes either through the upper or the lower slit, etc. That is to say: there are no superpositions of anything in nature. Superpositions concern only the wave function and its dynamics according to the Schrödinger equation (axiom (3)), but not the matter -- the objects -- that exist in the world, although, of course, the superpositions in the wave function and its dynamics can be relevant for the explanation of trajectories of the matter in physical space.

Axiom (2) then provides the particle dynamics. As is evident from the guiding equation, the evolution of the position of any particle depends, strictly speaking, on the position of all the other particles in the universe via the wave function. This is the manner in which BM implements quantum non-locality: it is correlated particle motion, with the correlation being established by the wave function and being independent of the distance of the particles. However, it is only correlated motion: by contrast to the collapse dynamics with the wave function representing an ontology of a wave in physical space in the guise of a matter density field (GRWm), there never is a delocation of matter in physical space. There always are only particles moving on continuous trajectories and thus with a finite velocity in physical space (in the relativistic setting, a velocity that is not greater than the velocity of light), with their motions being correlated with one another as a primitive matter of fact. The particle ontology and this dynamics provide a considerable advantage of the Bohmian account of measurement results over the collapse one: the particles are always there in space, instead of spontaneously localizing upon the collapse of the wave function -- either from nowhere, as the flashes do in GRWf, or being delocated all over space, as the matter density does in GRWm.

The ontology of BM -- known as the primitive ontology, that is, the referent of the formalism -- is given by the particles and their relative distances as well as the change of these distances (i.e. the particle motion). The wave function is defined by its dynamical role of providing the velocity field along which the particles move. The wave function is set out on configuration space. It can be conceived as a wave or a field; but, then, it is a wave or field on configuration space by contrast to an entity in physical space over and above the particles: the wave function does not have values at the points of physical space. It is therefore misleading to consider BM as an ontology of a dualism of particles and a wave and to imagine the wave function as a pilot wave that guides or pilotes the particles in physical space. These are metaphorical ways of speaking, since the wave function cannot be or represent a wave in physical space.

The wave function is nomological in the sense that it is introduced and defined through its dynamical role for the particle motion. Consequent upon its being nomological in that sense, all the stances in the metaphysics of laws of nature are applicable to the wave function. In particular, in recent years, a Bohmian Humeanism has been developed according to which the universal wave function is fixed by or supervenes on the history of the particle positions, being a variable that figures in the Humean best system (see \cite{Miller:2014aa}, \cite{Esfeld:2014aa}, \cite{Callender:2014aa} and \cite{Bhogal:2015aa}). This stance is applicable to all the theories that introduce the wave function through its dynamical role for the evolution of the configuration of matter in physical space, including GRWm and GRWf (see \cite{Dowker:2005aa} for a precise physical model based on GRWf).

This Humean stance makes clear that one can be a scientific realist without subscribing to an ontological commitment to the wave function and without falling into instrumentalism about the wave function: the role of the wave function is in the first place a dynamical one through its position in the law of motion (axiom (2)), with its role for the calculation of measurement outcome statistics deriving from that nomological role (axiom (4)). But the laws, including the dynamical parameters that figure in them, can be the axioms of the system that achieves the best balance between simplicity and information in representing the particle motion, as on Humeanism, instead of being entities that exist in the physical world over and above the configuration of matter (see \citet[][ch. 2.3]{Esfeld:2017aa}).       

The particle positions are an additional parameter in BM in the sense that the wave function and its evolution according to the Schrödinger equation do not contain the information about the exact particle positions and their evolution. All there is to the Bohmian particles are their relative positions, that is, the distances among them. Although the parameter of particle mass figures in the guiding equation \eqref{eq:standard-guiding-equation}, mass cannot be considered as an intrinsic property of the particles in BM. It is not situated where the particles are, but in the -- superposed -- wave packets. The same holds for the charge (see e.g. \cite{Brown:1995aa, Brown:1996aa} and references therein; cf. most recently \cite{Pylkkanen:2014aa} and \cite{Esfeld2014}).

The particle positions are the only additional parameter. Theorems like the one of \cite{Kochen:1967aa} prove that it is not possible to take the operators or observables of quantum mechanics to have definite values independently of measurement contexts on pain of violating the predictions of quantum mechanics for measurement outcome statistics. Nonetheless, these theorems leave the possibility open to admit one additional parameter of the quantum objects that has a definite value, without the precise information about that value figuring in the wave function. The natural choice for the additional parameter then is position, since all measurement outcomes consist in -- macroscopically recorded -- positions of discrete objects. In making this choice, one therefore lays the ground for solving the measurement problem. By contrast, pursuing a strategy that accords definite values to different parameters over time -- as done in so-called modal interpretations apart from BM -- falls victim to the measurement problem, as has been proven by \citet[][pp. 13-14]{Maudlin:1995aa}.

It is often maintained that apart from position, all the other operators or observables are treated as contextual properties in BM, in the sense that they acquire a definite value, signifying that they are realized as properties of quantum systems, only in the context of measurements. But this is wrong-headed. The operators or observables defined on a Hilbert space never are properties of anything. Suggesting that measurements somehow bring into existence properties that are contextual in the sense that they do not exist independently of measurement situations is a confused manner of talking. The operators or observables defined on a Hilbert space are instruments that provide information about how the quantum objects behave in certain situations, and that comes finally down to information about how their positions develop. Thus, there is no property of spin that quantum objects possess, and there is no contextual property of a definite value of spin in a certain direction that quantum objects acquire in the context of measurement situations. What these measurements do essentially is to provide information about particle positions: for instance, the measurement result ``spin up'' of the measurement of an electron provides the information that the particle is situated in the upwards wave packet, and not in the downwards one, etc. (see \citet[][ch. 4]{Bell:2004aa}, originally published 1971, and \cite{Norsen:2014aa}). A similar remark applies to the matter density field in GRWm and the flashes in GRWf -- again, they only have position, and the operators or observables provide information about their position (i.e. the position of flashes or the matter density at points of space).

Consequently, although if one formulates the relationship of the operators or observables defined on a Hilbert space in first order logic, the result is that this relationship does not fulfill the laws of classical logic, there is no reason to assume that the world is not conform to the laws of classical logic, on pain of confusing operators with properties of objects in the world. BM in particular shows why there is no problem with classical logic in quantum mechanics. More precisely and more generally speaking, when it comes to the ontology of what there is in the world and the dynamics of these entities, one solves the measurement problem by sticking to classical logic, and one would remain trapped by this problem if one were to abandon classical logic in the formulation of an ontology of quantum physics.

The upshot of these considerations is this one: in classical physics, one can be liberal about the properties of objects. That is to say, one can take the parameters that figure in the dynamical equations of classical physical theories to designate properties of the objects that these theories admit, such as the particles. Quantum physics teaches us that such a liberal ontological attitude is misplaced: the parameters that one employs to describe the dynamics of the physical objects -- including in particular their behaviour in measurement situations -- cannot without further reflection be attributed as properties to the objects. This holds even for the classical parameters of mass and charge, which are, in quantum physics, situated at the level of the wave function and can hence not be considered as intrinsic properties of the objects, although their value remains constant.

Quantum physics thereby teaches us that it is mandatory to draw the following distinction: on the one hand, there is the basic or primitive ontology of the theory, namely the referents of the formalism, which are supposed to be simply there in nature -- such as particles, which are then only characterized by what is necessary and minimally sufficient for something to be a particle, namely relative positions (the same holds for the flashes in GRWf and the positions of matter density values in GRWm). On the other hand, there is the dynamical structure of the theory. All and only the parameters that are introduced in terms of the functional or causal role that they exercise for the evolution of the elements of the basic or primitive ontology belong to the dynamical structure. They are nomological in the sense that they are there to perform a certain role for the evolution of the referents of the theory through their place in the laws of that evolution (but, of course, they are not themselves laws -- not even the universal wave function in BM is a law, since the theory admits of models with different universal wave functions). It then depends on the stance that one takes with respect to laws whether or not one accords the dynamical structure a place in the ontology over and above the basic or primitive ontology or takes it merely to be a means of representation (as on Humeanism).

Not only the operators in quantum mechanics, but also the classical parameters of mass and charge belong to the dynamical structure. Mass and charge are also in classical mechanics introduced for through their functional role for the particle motion (cf. \citet[][p. 241]{Mach:1919aa} on mass in Newtonian mechanics). Consequently, also in classical mechanics, there is the distinction available between on the one hand the primitive ontology of the theory in the guise of particle positions and particle motion and on the other hand the dynamical structure in terms of mass and charge, forces and fields, energy and potentials introduced through their causal role for the particle motion. Thus, also in classical mechanics, one can suspend or refuse an ontological commitment to the dynamical structure, being committed only to the particle positions and their change (see \citet[][§ 5.2]{Hall:2009aa} and \citet[][ch. 2.3]{Esfeld:2017aa}). In short, quantum physics makes simply evident that it is mandatory to draw a distinction that was already there all the time.

Although the particle positions make up the ontology in BM, there are limits to their accessibility. These limits are given in axiom (4), which implies that BM cannot make more precise predictions about measurement outcomes on sub-systems of the universe than those generated by using Born's rule. The link between the dynamical laws of BM and the $|\Psi|^2$--density on the level of the universal wave function as typicality measure is at least as tight as the link between the dynamical laws of classical mechanics in the Hamiltonian formulation and the Lebesgue measure (see \cite{Goldstein:2007aa}). Indeed, in BM, the quantum probabilities have the same status as the probabilities in classical statistical mechanics: they enter into the theory as the answer to the question of what evolution of a given system we can typically expect in situations in which the evolution of the system is highly sensitive to slight variations of its initial conditions and we do not know the exact initial conditions. In such situations, the deterministic laws of motion cannot be employed to generate deterministic predictions. Nonetheless, the probabilities are objective: they capture patterns in the evolution of the objects in the universe that show up when one considers many situations of the same type, such as many coin tosses in classical physics or running the double slit experiment with many particles in quantum physics. To put it in a nutshell, the Bohmian universe is like a classical universe in which not the motion of the planets, but the coin toss is the standard situation.

However, one may wonder whether if one lived in a classical universe of coin tosses, one would endorse the Hamiltonian laws as providing the dynamics of that universe -- Hamiltonian mechanics could then well be a minority position like BM today. The reason for endorsing Hamiltonian mechanics in that classical case would be the same as the one for endorsing BM in the quantum case: we need not only statistical predictions, but also a dynamics that describes the individual processes occurring in nature, on pain of falling into a measurement problem in the guise of the inability to account for the occurrence of determinate measurement outcomes. That is to say: in a classical universe of coin tosses, the Hamiltonian laws would be useless for predictions, as the calculation of Bohmian particle trajectories is useless for predictions. But both are indispensable for physics as a theory of nature. This is not only -- philosophical -- ontology; it is the business of physics to provide dynamical laws that apply to the individual processes in nature.        

Nonetheless, in contrast to the relationship between classical statistical mechanics and classical mechanics, there is a principled limit to the accessibility of initial conditions of physical systems in the quantum case. That limit becomes evident, for instance, in the famous Heisenberg uncertainty relations. It is trivial that measurement is an interaction so that the measurement changes the measured system and can thus not simply reveal the position and velocity that it had independently of the measurement interaction. Some limit to the accessibility of physical systems may follow from the triviality that any measurement is an interaction. Thus, it is well known in classical physics that no observer within the universe could obtain the data that Laplace's demon would need for its predictions. However, the quantum case is not simply an illustration of that triviality, since there is a precise principled limit of the epistemic accessibility of quantum systems (as illustrated, for instance, by the Heisenberg uncertainty relations). In BM, this principled limit follows from applying the theory as defined above to measurement interactions (see \citet[][ch. 2]{Durr:2013aa}).

In the last resort, of course, it is the particle motion in the world that makes possible stable particle correlations such that one particle configuration (say a measurement device, or a brain) records the position and traces the motion of other particles and particle configurations, and it is also the particle motion in the world that puts a limit on such correlations. The laws of BM including the typicality measure and the assumption that the actual universe is a typical Bohmian universe bring out these facts about the actual particle motion. Again, the wave function in configuration space represents that particle motion; it is not the wave function that puts the limit on the epistemic accessibility of the particle positions, although we understand that limit by representing the particle motion through the wave function. Instead of taking this limit to be a drawback and hoping for a physical theory like classical mechanics in which there are paradigmatic cases of deterministic laws of motion enabling deterministic predictions (e.g. the motion of the planets), it is fortunate and by no means trivial that there are such stable particle correlations in the universe at all so that we can represent actual particle positions and motions and make reliable predictions.      

As these considerations make clear, posing a limit to the accessibility of the objects in physical space is by no means a feature that is peculiar to BM. Such a limit applies not only to the particles in BM, but, for instance, also to the flash distribution in GRWf and the matter density field in GRWm (see \cite{Cowan:2015aa}). In any case of a quantum ontology of objects in physical space, if these objects were fully accessible, we could the thus gained information employ to exploit quantum non-locality for superluminal signaling. This limited accessibility of the particle configuration, the flash distribution or the matter density field confirms that if one endorses proposition (3.A and B) of the measurement problem -- i.e. determinate measurement outcomes whose statistical distributions are given by Born's rule --, the situation is not that one has to reject \emph{either} proposition (1.A and B) \emph{or} proposition (2.A and B). One has in this case to modify \emph{both} proposition (1.A and B) and proposition (2.A and B). If one starts starts from admitting particle positions that are not revealed by the wave function (rejection of proposition (1. A and B)), one can retain the Schrödinger dynamics for the wave function (proposition (2.A and B), but this then is not the complete dynamics: the central dynamical law then is the law of the evolution of the additional variables, namely the guiding equation that tells us how the particle positions evolve in physical space. If one starts from amending the Schrödinger equation by collapse parameters (rejection of proposition (2. A and B)), one can retain the wave function and its dynamics as describing the evolution of the objects in physical space (flashes, matter density field) (proposition (1.A and B), but that distribution then nevertheless is ``hidden'' in the sense that it follows from the theory that it is not fully accessible.

At least three conclusions can be drawn from this situation:
\begin{enumerate}
\item Any theory that admits definite measurement outcomes distinguishes position -- be it particle positions, positions of flashes, or values of matter density at points of space. All the other observables are accounted for on this basis.
\item If one is not prepared to accept a principled limit to the epistemic accessibility of the objects in physical space, one remains trapped by the measurement problem, because one then does not have a dynamics at one's disposal that accounts for determinate measurement outcomes.
\item The solution space to the measurement problem reduces to this one: \emph{either} one abandons determinate measurement outcomes in physical space, in which case one can retain the propositions of the wave function being complete and its evolving always according to the Schrödinger equation; one then has to come up with an Everett-style account of why it appears to us as if there were determinate measurement outcomes in physical space. \emph{Or} one retains determinate measurement outcomes in physical space, and the account of these measurement outcomes then commits one to endorsing a distribution of objects in physical space whose evolution cannot be given by the Schrödinger equation and is not fully accessible.   
\end{enumerate}

\noindent BM can be seen as the answer to the following question: What is the minimal deviation from classical mechanics that is necessary in order to obtain quantum mechanics? BM shows that the physical ontology can remain the same -- point particles moving on continuous trajectories -- and that the status of probabilities can remain unchanged. What has to change is the dynamics, that is, a wave function parameter has to be introduced with the wave function binding the evolution of the particle positions together independently of their distance in physical space.

That notwithstanding, this conceptualisation of quantum non-locality is like Newtonian gravitation in that there never is any matter instantaneously delocated in space: as in Newtonian gravitation the distribution of the particle positions, velocities and masses all over space at any time $t$ fixes the acceleration of the particles at that $t$, so in BM the distribution of the particle positions and the universal wave function at any time $t$ fix the velocity of the particles at that $t$. Of course, Newtonian gravitation concerns all particles indiscriminately and depends on the square of their distance, whereas quantum non-locality is \emph{de facto} highly selective -- i.e. concerns \emph{de facto} only specific particles -- and is independent of their distance. Nonetheless, in both cases, non-locality means that there are correlations in the particle motion without these correlations being mediated by the -- instantaneous -- transport of anything all over space. There is no reason to change more. Doing so only leads to unpalatable consequences beyond the quantum non-locality with which one has to come to terms anyway.

\section{Permanent particles in quantum field theory}
The same conclusions apply to quantum field theory (QFT). Let us briefly point out why (for details, see \citet[][ch. 4.2]{Esfeld:2017aa}). The measurement problem hits QFT in the same way as quantum mechanics (see \cite{Barrett:2014aa}). Again, we have a highly successful formalism to calculate measurement outcome statistics at our disposal. However, the measurement problem as formulated by \cite{Maudlin:1995aa} arises as soon as it comes to accommodating measurement outcomes in physical space, for this formalism does as such not include a dynamics that describes the individual processes in nature that lead to determinate measurement outcomes. In particular, only the formalism to calculate measurement outcome statistics is Lorentz-invariant (i.e. it is irrelevant for it in which temporal order space-like separated measurement outcomes occur). But we do not have a relativistic, Lorentz-invariant dynamics of the individual processes at our disposal that lead to determinate measurement outcomes in physical space (despite what may look like claims to the contrary in the context of GRWf and GRWm; see \cite{Tumulka:2006aa} and \cite{Bedingham:2014aa} for these claims; see \cite{Barrett:2014aa} and \cite{Esfeld:2014ad} for pointing out their limits).

Despite its name, QFT is not an ontology of fields. The fields in the formalism are operator valued fields, by contrast to fields that have determinate values at the points of physical space. In the standard model of particle physics, fields are there to model the interactions (i.e. the electromagnetic, the weak and the strong interaction, without gravitation). If one endorses an ontology of fields in physical space, then the problem is, like in GRWm for quantum mechanics, to formulate a credible dynamics of the contraction of fields so that they can constitute determinate measurement outcomes and, in general, discrete macroscopic objects.

However, also a particle ontology faces obstacles in QFT, namely new obstacles that do not arise in quantum mechanics: in QFT, measurement records not only fail to keep track of particle trajectories, but, moreover, they fail to keep track of a fixed number of particles. Also in QFT, as in any other area of physics, the experimental evidence recorded by the measurement devices is particle evidence. But this evidence includes what appears to be particle creation and annihilation events, so that there seems to be no fixed number of particles that persist.

Nonetheless, these experimental facts do as such not entitle any inferences for ontology. More precisely, as it is a \emph{non sequitur} to take particle trajectories to be ruled out in quantum mechanics due to the Heisenberg uncertainty relations, so it is a \emph{non sequitur} to take permanent particles moving on definite trajectories according to a deterministic law to be ruled out in QFT due to the statistics of particle creation and annihilation phenomena. In both cases, the experimental evidence leaves open whether the particle trajectories do not exist or are simply not accessible in measurements and whether the particles come into being and are annihilated or whether it is simply not possible to keep track of them in the experiments. These issues have to be settled by the theory. The standard for assessing the theory is the solution to the measurement problem. 

It is possible to pursue a Bohmian solution to the measurement problem in QFT along the same lines as in quantum mechanics. As BM has no ambition to improve on the statistical predictions of measurement outcomes in quantum mechanics, but deduce these predictions from the axiom that the universe is in quantum equilibrium, so there is no ambition that a Bohmian solution to the measurement problem in QFT can resolve the mathematical difficulties that QFT currently faces. That is to say: Bohmian QFT has to rely on cut-offs as does the standard model of particle physics when it comes to dynamical laws of interactions (by contrast to scattering theory). Given appropriate cut-offs, one can formulate a Bohmian theory for QFT in the same way as for quantum mechanics: on what is known as Bohmian Dirac sea QFT, the ontology is one of a very large, but finite and fixed number of permanent point particles that move on continuous trajectories as given by a deterministic dynamical law (guiding equation) by means of the universal wave function.

More precisely, one can define a ground state for these particles that is a state of equilibrium. This state is one of a homogeneous particle motion in the sense that the particle interactions cancel each other out. Consequently, the particle motion is not accessible. This state corresponds to what is known as the vacuum state. However, on this view, it is not at all a vacuum, but a sea full of particles (known as the Dirac sea) in which the particles are not accessible. What is accessible and what is effectively modeled by the Fock space formalism of calculating measurement outcome statistics are the excitations of this ground state that show up in what appears to be particle creation and annihilation events. Again, by defining a typicality measure on the level of the universal wave function, one can derive the predictions of measurement outcome statistics in the guise of, in this case, statistics of excitation events from the ground state. Thus, again, the quantum probabilities are due to a -- principled -- limit to the accessibility of the particle motion (see \cite{Colin:2007aa} and \citet[][ch. 4.2]{Esfeld:2017aa}).            

When pursuing a solution to the measurement problem in terms of an ontology of particles, it is worthwhile to go down all the Bohmian way also in QFT. What is known as Bell-type Bohmian QFT (\citet[][ch. 19]{Bell:2004aa} and further elaborated on in \cite{durr_bell-type_2005}) goes only half the way down: on this theory, the particles come into and go out of existence with statistical jumps between sectors of different particle numbers in the dynamics. However, this proposal amounts to elevating what is known as quasi particles that are dependent on the -- contingent -- choice of a reference frame to the status of particles in the ontology. Furthermore, it is committed to absolute space as the substance in which the particles come into and go out of existence. The Bell-type (quasi)particles are much like the GRWf flashes, apart from the fact that they can persist for a limited time, instead of being ephemeral. By contrast, on the Bohmian Dirac sea theory, the particles are permanent so that they can be conceived as being individuated by the distances among them in any given configuration and as having an identity in change through the continuous trajectories that their motion traces out. Consequently, there is no need for a commitment to a surplus structure in the guise of absolute space and time in the ontology; there is no need for a medium in which the particles exist (viz. come into and go out of existence). Probabilities then come in through linking the deterministic dynamics with a typicality measure. Filling negative energy states with particles is no problem on this ontology, since the only property of the particles is their position; energy is not a property of anything, but a variable in the formalism to track particle motion.

In sum, on the Bohmian Dirac sea ontology, the account of measurement outcomes is of the same type as in Bohmian quantum mechanics, with the difference that there are much more particles in the sea than one would expect in an ontology of particle positions only that are given by the distances among the particles (corresponding to empty space in the representation in terms of a space in which the particles are embedded). The account again has two stages: the ontology of particles -- in this case, the excitations of particles against the background of the particle motion in the Dirac sea -- accounts for the presence of the measured quantum objects as well as the one of the macroscopic systems; the latter are constituted by these particle excitations, with the particle dynamics that yields these excitations explaining their stability. The measurement outcome statistics then are accounted for in terms of the limited accessibility of the quantum particles by means of defining a typicality measure from which one then deduces the formalism to calculate these statistics.        

In any case, the objects that one poses in an ontology of quantum physics are theoretical entities. They are admitted to explain the phenomena as given by the measurement outcome statistics. That is why the solution to the measurement problem is the standard for assessing these proposals. In any case, if one admits quantum objects in physical space beyond the wave function, there is a limit to their accessibility; the wave function has in this case exclusively a dynamical status, namely yielding the dynamics for these objects. The Bohmian solution to the measurement problem provides the least deviation from the ontology of classical mechanics that is necessary to accommodate quantum physics, both in the case of quantum mechanics and in the case of quantum field theory. There is no cogent reason to go beyond that minimum.

\bibliographystyle{apalike}
\bibliography{references_fundont}

\begin{thebibliography}{}

\bibitem[Albert, 2015]{Albert:2015aa}
Albert, D.~Z. (2015).
\newblock {\em After physics}.
\newblock Cambridge, Massachusetts: Harvard University Press.

\bibitem[Allori et~al., 2008]{Allori:2008aa}
Allori, V., Goldstein, S., Tumulka, R., and Zangh{\`\i}, N. (2008).
\newblock On the common structure of {B}ohmian mechanics and the
  {G}hirardi-{R}imini-{W}eber theory.
\newblock {\em British Journal for the Philosophy of Science}, 59(3):353--389.

\bibitem[Allori et~al., 2014]{Allori:2013aa}
Allori, V., Goldstein, S., Tumulka, R., and Zangh{\`\i}, N. (2014).
\newblock Predictions and primitive ontology in quantum foundations: a study of
  examples.
\newblock {\em British Journal for the Philosophy of Science}, 65(2):323--352.

\bibitem[Barrett, 2014]{Barrett:2014aa}
Barrett, J.~A. (2014).
\newblock Entanglement and disentanglement in relativistic quantum mechanics.
\newblock {\em Studies in History and Philosophy of Modern Physics},
  48:168--174.

\bibitem[Bedingham et~al., 2014]{Bedingham:2014aa}
Bedingham, D., D\"{u}rr, D., Ghirardi, G.~C., Goldstein, S., Tumulka, R., and
  Zangh{\`\i}, N. (2014).
\newblock Matter density and relativistic models of wave function collapse.
\newblock {\em Journal of Statistical Physics}, 154:623--631.

\bibitem[Bell, 2004]{Bell:2004aa}
Bell, J.~S. (2004).
\newblock {\em Speakable and unspeakable in quantum mechanics}.
\newblock Cambridge: Cambridge University Press, second edition.

\bibitem[Bhogal and Perry, 2017]{Bhogal:2015aa}
Bhogal, H. and Perry, Z.~R. (2017).
\newblock What the {H}umean should say about entanglement.
\newblock {\em No{\^u}s}, 51(1):74--94.

\bibitem[Bohm, 1952]{Bohm:1952}
Bohm, D. (1952).
\newblock A suggested interpretation of the quantum theory in terms of
  ``hidden'' variables.
\newblock {\em Physical Review}, 85(2):166--179, 180--193.

\bibitem[Brown et~al., 1995]{Brown:1995aa}
Brown, H.~R., Dewdney, C., and Horton, G. (1995).
\newblock Bohm particles and their detection in the light of neutron
  interferometry.
\newblock {\em Foundations of Physics}, 25(2):329--347.

\bibitem[Brown et~al., 1996]{Brown:1996aa}
Brown, H.~R., Elby, A., and Weingard, R. (1996).
\newblock Cause and effect in the pilot-wave interpretation of quantum
  mechanics.
\newblock In Cushing, J.~T., Fine, A., and Goldstein, S., editors, {\em Bohmian
  mechanics and quantum theory: an appraisal}, volume 184 of {\em Boston
  Studies in the Philosophy of Science}, pages 309--319. Dordrecht: Springer.

\bibitem[Callender, 2015]{Callender:2014aa}
Callender, C. (2015).
\newblock One world, one beable.
\newblock {\em Synthese}, 192(10):3153--3177.

\bibitem[Colin and Struyve, 2007]{Colin:2007aa}
Colin, S. and Struyve, W. (2007).
\newblock A {D}irac sea pilot-wave model for quantum field theory.
\newblock {\em Journal of Physics A}, 40(26):7309--7341.

\bibitem[Cowan and Tumulka, 2016]{Cowan:2015aa}
Cowan, C.~W. and Tumulka, R. (2016).
\newblock Epistemology of wave function collapse in quantum physics.
\newblock {\em British Journal for the Philosophy of Science}, 67:405--434.

\bibitem[Curceanu and alteri, 2016]{Curceanu:2016aa}
Curceanu, C. and alteri (2016).
\newblock Spontaneously emitted x-rays: an experimental signature of the
  dynamical reduction models.
\newblock {\em Foundations of Physics}, 46(3):263--268.

\bibitem[de~Broglie, 1928]{Broglie:1928aa}
de~Broglie, L. (1928).
\newblock La nouvelle dynamique des quanta.
\newblock {\em Electrons et photons. Rapports et discussions du cinqui{\`e}me
  Conseil de physique tenu {\`a} Bruxelles du 24 au 29 octobre 1927 sous les
  auspices de l'Institut international de physique Solvay}, pages 105--132.
\newblock Paris: Gauthier-Villars. English translation in Bacciagaluppi, G. and
  Valentini, A., editors (2009). \emph{Quantum theory at the crossroads.
  Reconsidering the 1927 Solvay conference}, pages 341--371. Cambridge:
  Cambridge University Press.

\bibitem[de~Broglie, 1964]{Broglie:1964aa}
de~Broglie, L. (1964).
\newblock {\em The current interpretation of wave mechanics. A critical study}.
\newblock Amsterdam: Elsevier.

\bibitem[Dowker and Herbauts, 2005]{Dowker:2005aa}
Dowker, F. and Herbauts, I. (2005).
\newblock The status of the wave function in dynamical collapse models.
\newblock {\em Foundations of Physics Letters}, 18:499--518.

\bibitem[D{\"u}rr et~al., 2005]{durr_bell-type_2005}
D{\"u}rr, D., Goldstein, S., Tumulka, R., and Zangh{\`\i}, N. (2005).
\newblock Bell-type quantum field theories.
\newblock {\em Journal of Physics A: Mathematical and General}, 38(4):R1--R43.

\bibitem[D{\"u}rr et~al., 2013]{Durr:2013aa}
D{\"u}rr, D., Goldstein, S., and Zangh{\`\i}, N. (2013).
\newblock {\em Quantum physics without quantum philosophy}.
\newblock Berlin: Springer.

\bibitem[Egg and Esfeld, 2014]{Egg:2014aa}
Egg, M. and Esfeld, M. (2014).
\newblock Non-local common cause explanations for {EPR}.
\newblock {\em European Journal for Philosophy of Science}, 4:181--196.

\bibitem[Egg and Esfeld, 2015]{Egg:2015aa}
Egg, M. and Esfeld, M. (2015).
\newblock Primitive ontology and quantum state in the {GRW} matter density
  theory.
\newblock {\em Synthese}, 192(10):3229--3245.

\bibitem[Esfeld, 2014]{Esfeld:2014aa}
Esfeld, M. (2014).
\newblock Quantum {H}umeanism, or: physicalism without properties.
\newblock {\em The Philosophical Quarterly}, 64(256):453--470.

\bibitem[Esfeld and Deckert, 2017]{Esfeld:2017aa}
Esfeld, M. and Deckert, D.-A. (2017).
\newblock {\em A minimalist ontology of the natural world}.
\newblock New York: Routledge.

\bibitem[Esfeld and Gisin, 2014]{Esfeld:2014ad}
Esfeld, M. and Gisin, N. (2014).
\newblock The {GRW} flash theory: a relativistic quantum ontology of matter in
  space-time?
\newblock {\em Philosophy of Science}, 81:248--264.

\bibitem[Esfeld et~al., 2017]{Esfeld2014}
Esfeld, M., Lazarovici, D., Lam, V., and Hubert, M. (2017).
\newblock The physics and metaphysics of primitive stuff.
\newblock {\em British Journal for the Philosophy of Science}, 68:133--161.

\bibitem[Ghirardi et~al., 1995]{Ghirardi:1995aa}
Ghirardi, G.~C., Grassi, R., and Benatti, F. (1995).
\newblock Describing the macroscopic world: closing the circle within the
  dynamical reduction program.
\newblock {\em Foundations of Physics}, 25(1):5--38.

\bibitem[Ghirardi et~al., 1990]{Ghirardi:1990aa}
Ghirardi, G.~C., Pearle, P., and Rimini, A. (1990).
\newblock Markov processes in {H}ilbert space and continuous spontaneous
  localization of systems of identical particles.
\newblock {\em Physical Review A}, 42:78--89.

\bibitem[Ghirardi et~al., 1986]{Ghirardi:1986aa}
Ghirardi, G.~C., Rimini, A., and Weber, T. (1986).
\newblock Unified dynamics for microscopic and macroscopic systems.
\newblock {\em Physical Review D}, 34(2):470--491.

\bibitem[Goldstein and Struyve, 2007]{Goldstein:2007aa}
Goldstein, S. and Struyve, W. (2007).
\newblock On the uniqueness of quantum equilibrium in {B}ohmian mechanics.
\newblock {\em Journal of Statistical Physics}, 128(5):1197--1209.

\bibitem[Hall, 2009]{Hall:2009aa}
Hall, N. (2009).
\newblock Humean reductionism about laws of nature.
\newblock Unpublished manuscript. http://philpapers.org/rec/halhra.

\bibitem[Kochen and Specker, 1967]{Kochen:1967aa}
Kochen, S. and Specker, E. (1967).
\newblock The problem of hidden variables in quantum mechanics.
\newblock {\em Journal of Mathematics and Mechanics}, 17:59--87.

\bibitem[Mach, 1919]{Mach:1919aa}
Mach, E. (1919).
\newblock {\em The science of mechanics: a critical and historical account of
  its development. Fourth edition. Translation by Thomas J. McCormack}.
\newblock Chicago: Open Court.

\bibitem[Maudlin, 1995]{Maudlin:1995aa}
Maudlin, T. (1995).
\newblock Three measurement problems.
\newblock {\em Topoi}, 14:7--15.

\bibitem[Maudlin, 2010]{Maudlin:2010aa}
Maudlin, T. (2010).
\newblock Can the world be only wave-function?
\newblock In Saunders, S., Barrett, J., Kent, A., and Wallace, D., editors,
  {\em Many worlds? Everett, quantum theory, and reality}, pages 121--143.
  Oxford: Oxford University Press.

\bibitem[Maudlin, 2011]{Maudlin:2011aa}
Maudlin, T. (2011).
\newblock {\em Quantum non-locality and relativity. Third edition}.
\newblock Chichester: Wiley-Blackwell.

\bibitem[Miller, 2014]{Miller:2014aa}
Miller, E. (2014).
\newblock Quantum entanglement, {B}ohmian mechanics, and {H}umean
  supervenience.
\newblock {\em Australasian Journal of Philosophy}, 92:567--583.

\bibitem[Norsen, 2005]{Norsen:2005aa}
Norsen, T. (2005).
\newblock Einstein's boxes.
\newblock {\em American Journal of Physics}, 73:164--176.

\bibitem[Norsen, 2014]{Norsen:2014aa}
Norsen, T. (2014).
\newblock The pilot-wave perspective on spin.
\newblock {\em American Journal of Physics}, 82(4):337--348.

\bibitem[Pylkk{\"a}nen et~al., 2015]{Pylkkanen:2014aa}
Pylkk{\"a}nen, P., Hiley, B.~J., and P{\"a}ttiniemi, I. (2015).
\newblock Bohm's approach and individuality.
\newblock In Guay, A. and Pradeu, T., editors, {\em Individuals across the
  sciences}, chapter~12, pages 226--246. Oxford: Oxford University Press.

\bibitem[Rovelli, 1997]{Rovelli:1997aa}
Rovelli, C. (1997).
\newblock Halfway through the woods: contemporary research on space and time.
\newblock In Earman, J. and Norton, J., editors, {\em The cosmos of science},
  pages 180--223. Pittsburgh: University of Pittsburgh Press.

\bibitem[Schr{\"o}dinger, 1935]{Schroedinger:1935aa}
Schr{\"o}dinger, E. (1935).
\newblock Die gegenw{\"a}rtige {S}ituation in der {Q}uantenmechanik.
\newblock {\em Naturwissenschaften}, 23:807--812.

\bibitem[Tumulka, 2006]{Tumulka:2006aa}
Tumulka, R. (2006).
\newblock A relativistic version of the {G}hirardi--{R}imini--{W}eber model.
\newblock {\em Journal of Statistical Physics}, 125(4):821--840.

\bibitem[Vassallo and Ip, 2016]{Vassallo:2016aa}
Vassallo, A. and Ip, P.~H. (2016).
\newblock On the conceptual issues surrounding the notion of relational
  {B}ohmian dynamics.
\newblock {\em Foundations of Physics}, 46(8):943--972.

\bibitem[Wallace, 2014]{Wallace:2014aa}
Wallace, D. (2014).
\newblock Life and death in the tails of the {GRW} wave function.
\newblock {\em arXiv:1407.4746 [quant-ph]}.

\end{thebibliography}

\end{document}